\documentclass[twocolumn,showpacs,preprintnumbers,superscriptaddress,amsmath,amssymb]{revtex4}
\usepackage{graphicx}
\usepackage{dcolumn}
\usepackage{bm}

\begin{document}

\preprint{APS/123-QED}

\title{Temperature dependent Luttinger surfaces}

\author{T. Ito}
\email[]{tito@ims.ac.jp}
\affiliation{The Institute of Physical and Chemical Research (RIKEN), Sayo-gun, Hyogo 679-5143, Japan}
\author{A. Chainani}
\affiliation{The Institute of Physical and Chemical Research (RIKEN), Sayo-gun, Hyogo 679-5143, Japan}
\author{T. Haruna}
\affiliation{Institute for Solid State Physics (ISSP), University of Tokyo, Kashiwa, Chiba 277-8581, Japan}
\author{K. Kanai}
\affiliation{The Institute of Physical and Chemical Research (RIKEN), Sayo-gun, Hyogo 679-5143, Japan}
\author{T. Yokoya}
\affiliation{Institute for Solid State Physics (ISSP), University of Tokyo, Kashiwa, Chiba 277-8581, Japan}
\author{S. Shin}
\affiliation{The Institute of Physical and Chemical Research (RIKEN), Sayo-gun, Hyogo 679-5143, Japan}
\affiliation{Institute for Solid State Physics (ISSP), University of Tokyo, Kashiwa, Chiba 277-8581, Japan}
\author{R. Kato}
\affiliation{The Institute of Physical and Chemical Research (RIKEN), Wako, Saitama 351-0198, Japan}

\date{\today}

\begin{abstract}
The Luttinger surface of an organic metal (TTF-TCNQ), 
possessing charge order and spin-charge separation, is investigated 
using temperature dependent angle-resolved photoemission spectroscopy.  
The Luttinger surface topology, obtained from momentum distribution curves, 
changes from quasi-2D(dimensional) to quasi-1D with temperature.  
The high temperature quasi-2D surface exhibits 4$k_F$ charge-density-wave (CDW) superstructure 
in the TCNQ derived holon band, in the absence of 2$k_F$ order.  
Decreasing temperature results in quasi-1D nested 2$k_F$ CDW order 
in the TCNQ spinon band and in the TTF surface.  
The results establish the link in momentum-space between charge order and spin-charge separation in a Luttinger liquid.
\end{abstract}
\pacs{71.20.-b, 71.10.-w, 79.60.-i}
\maketitle
The Fermi surface defining a metal \cite{Ashcroft} is a set of points in momentum ($k$) space 
at which a dispersing quasiparticle band crosses the Fermi-level ($E_F$) in energy distribution curves (EDCs), 
with corresponding peaks in momentum distribution curves (MDCs) \cite{Valla}.  
This holds for a correlated Fermi liquid. In contrast, many Luttinger liquids do not show 
quasiparticle band-crossings at $E_F$ in EDCs as measured by angle-resolved photoemission spectroscopy (ARPES) \cite{Allen}.  
This is also true for many quasi-2D high-$T_c$ cuprates and related materials \cite{Ren,Kim,Orgad,Gweon,Kivelson}.  
For the cuprates, this has been shown to be due to electron fractionalization or spin-charge separation in a quasi-1D Luttinger liquid.  
Recent theory \cite{Dzyaloshinskii} has also predicted that Luttinger liquids 
should exhibit a so-called `Luttinger surface' - the equivalent of a Fermi surface for a non-Fermi liquid.  

TTF - TCNQ ( tetrathiafulvalene - tetracyanoquinodimethane ), 
is an organic metal consisting of TTF (donor) and TCNQ (acceptor) columns stacked along the b-axis, 
and has been extensively studied for its structural, transport and thermodynamic properties \cite{Heeger,Tomkiewicz,Pouget,Shirane,Kagoshima,Gruner}.  
It is a classic example of a quasi 1-D correlated metal with a Peierls' transition, 
exhibiting 2$k_F$ and 4$k_F$ CDW order along the b* (or $k_y$) direction 
in $k$-space \cite{Heeger,Tomkiewicz,Pouget,Shirane,Kagoshima,Gruner}.  
These two aspects, namely (i) the strong electron-electron correlations and 
(ii) the electron-phonon coupling induced Peierls' transition setting in at 54~K, 
are believed to occur in different parts of the electronic structure : 
the TTF electronic structure displays the 2$k_F$ Peierls' order 
and the TCNQ part exhibits the 4$k_F$ Coulomb correlation driven order \cite{Heeger,Tomkiewicz,Gruner,Jerome,Zwick}.  
More recently, this system has also served as a very good example of a correlated metal 
exhibiting spin-charge separated branched dispersions of the 1-D Hubbard model, 
thus confirming its non-Fermi liquid character \cite{Claessen,Carmelo}.  
In this work, we investigate the relation between spin-charge separated dispersions 
and the temperature-evolution of 2$k_F$ and 4$k_F$ CDW order involving specific electronic states 
constituting the TTF or TCNQ Luttinger surfaces. 

Angle-resolved photoemission spectra were obtained from single crystals of TTF-TCNQ 
using a Scienta SES 2002 analyzer and a Gammadata-discharge lamp (He II$\alpha$, h$\nu$~=~40.8~eV).  
The energy and momentum resolutions for the ARPES measurments were 20~meV and 
0.01 \AA$^{-1}$, respectively.  
A thin film filter \cite{Yokoya} was instrumental in obtaining the spectra 
without degradation as a function of time and exposure to UV photons over several hours.  
The samples were cooled using a flowing liquid He cryostat and the temperature was controlled 
within $\pm$2~K using a temperature controller.  
Fresh surfaces were obtained by cleaving the samples in-situ 
and the temperature dependence was confirmed to be reproducible.

In Fig. 1(a)-(c) we plot two dimensional intensity maps 
as a function of momentum in the first Brillouin zone of TTF-TCNQ, 
for T = 200~K, 100~K and 30~K, obtained from ARPES spectra.  
The intensity maps are obtained for an energy window of 50~meV ($E_F$ to 50~meV binding energy) 
and the high intensity regions correspond to Luttinger surfaces in the first Brillouin zone.  
At 200~K, the main Luttinger surfaces consisting of the TCNQ surface 
(along slice 1, centered at $k_y$ $\sim$ $\pm$ 0.17b*) 
and the TTF surface (along slice 2, centered at $k_y$ $\sim$ $\pm$ 0.13b*), 
are well-separated in $k$-space and quasi-2D.  These are shown schematically 
in Fig. 1(e) as red (TCNQ) and yellow (TTF) regions.  
Additional replica features only for the TCNQ branch are seen in Fig. 1(a) 
at momenta translated by a value of $\sim$ 0.6b*, 
along slice 1 (brown regions in schematic Fig. 1(e)).  
The observation of replica features translated by $\sim$ 0.6b* correspond to the observation of 4$k_F$ order, 
consistent with superstructure spots in structural studies \cite{Pouget,Shirane,Kagoshima}.  
This surface arises from the holon band crossing at 4$k_F$, 
which is absent in band structure calculations \cite{Ishibashi}, 
but is obtained in the 1-D Hubbard model \cite{Claessen,Carmelo}.  
While the high energy band dispersions have been reported for the TCNQ holon band \cite{Claessen}, 
a Luttinger surface obtained from low energy 
\linebreak
 \begin{figure}
 \includegraphics[scale=.70]{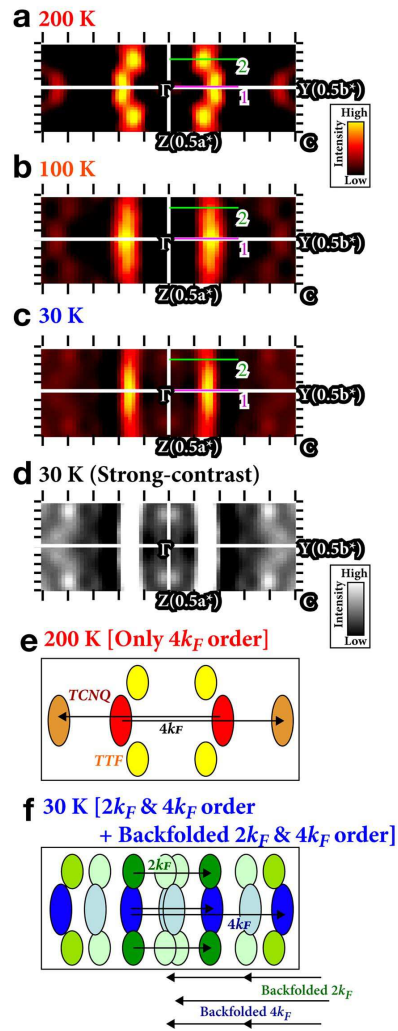}%
 \caption{\label{fig1}
 (Color) Temperature dependent Luttinger surface maps: 
 Fig. 1 (a-c) show the Luttinger surfaces of TTF-TCNQ as a function of temperature (200-100-30~K), 
 obtained from angle-resolved photoemission spectra.  
 Fig. 1(a) shows the TTF (along slice 1, centered at $k_y$ $\sim$ $\pm$ 0.13b*) 
 and TCNQ (along slice 1, centered at $k_y$ $\sim$ $\pm$ 0.17b*) main Luttinger surfaces 
 are separated in momentum space (see also schematic Fig. 1(e)).  
 The TCNQ surface shows a 4$k_F$ CDW superstructure at about $\pm$ 0.43b* along slice 1 ($\Gamma$-Y) in $k$-space.  
 The T = 100~K map (Fig.~1(b)) show features due to the 2$k_F$ CDW order in addition to the 4$k_F$ CDW vectors 
 which evolve with reducing temperature.  
 Fig. 1(c) shows that the low temperature (T = 30~K) main TTF and TCNQ Luttinger surfaces are nearly 1-D, 
 in good agreement with a nesting picture of the 2$k_F$ CDW vector.  
 The main Luttinger surfaces occur at $k_y$ $\sim$ $\pm$ 0.15b* and the 4$k_F$ superstructures are 
 at $\sim$ $\pm$ 0.45b* along slice 1 ($\Gamma$-Y) in $k$-space (dark blue and green regions shown schematically in Fig.~1(f) ).  
 Fig.~1(d) is a gray scale map of Fig.~1(c), showing the weaker backfolded structures 
 (at $k_y$ $\sim$ 0b* and $\pm$ 0.3b*) more clearly.  
 Along slice 2, the superstructures reflect deviations from perfect nesting due to small differences 
 in peak positions and widths of the main Luttinger surfaces.  Light blue and geeen regions 
 in schematic Fig.~1(f) show the relation of the backfolded features with the 2$k_F$ and 4$k_F$ CDW order.
 }
 \end{figure}
\noindent
spectral weight (within 50~meV of $E_F$) 
has not been possible earlier 
due to the very low intensity near $E_F$.  
The holon band derived Luttinger surface confirms the validity of the 1-D Hubbard model at low energies.
It's presence also indicates that the main TCNQ band along slice 1 is the spinon band 
and not simply the uncorrelated TCNQ band expected from band structure calculations.  
We have measured and confirmed the high energy dispersions (not shown) to be 
in excellent agreement with earlier work \cite{Zwick,Claessen}.  
At T = 200~K, since the separation between the main TTF and TCNQ bands are not equal to 2$k_F$, 
the 4$k_F$ order exists in the absence of 2$k_F$ order.  
The 2$k_F$ CDW order is expected to show up only below 150~K, 
as is known from earlier studies \cite{Pouget,Shirane,Kagoshima}.

At T = 100~K, the separation along the b* direction seen 
in the TTF and TCNQ branches is reduced and they merge together (Fig. 1(b)), 
with the observation of additional weak features in the map.  
These changes occur above the Peierls' transition (which sets in below 54~K), 
and correspond to the growth of the 2$k_F$ CDW order.  
While these weak features derived from the low energy electronic structure were not observed 
in earlier ARPES studies \cite{Zwick,Claessen}, 
they are consistent with a recent analysis of X-ray diffuse scattering above the Peierls' transition, 
due to quasi 1-D structural fluctuations \cite{Pouget04}.  
At the lowest temperature of 30~K (Fig. 1(c)), below the Peierls' transition at 54~K followed by additional order at 38~K, 
the TTF and TCNQ parts of the main Luttinger surfaces satisfy the nesting condition for the 2$k_F$ CDW order and are nearly 1-D.

MDCs obtained at T = 30~K with high signal to noise ratio, using the same energy window of 50~meV as for the maps of Fig. 1, 
are shown along $\Gamma$-Y (Fig. 2(a), slice 1) and along Z-C (Fig. 2(b), slice 2).  
The MDC curves show sharp peaks at equidistant points in $k$-space 
from the $\Gamma$ (Fig. 2(a)) and Z-points (Fig. 2(b)) at a value of $\sim$ ($\pm$ 0.15 $\pm$ 0.01)b*.  
The peak to peak separation in $k$-space($\sim$ 0.30b*) matches the CDW vector (2$k_F$ = 0.295b*) in TTF-TCNQ, 
known from structural studies \cite{Pouget,Shirane,Kagoshima}.  
Similar peaks in MDCs have been observed 
in the quasi 1-D conductors K$_{0.3}$MoO$_3$ \cite{Fedorov} and (TaSe$_4$)$_2$I \cite{Perfetti}, 
which undergo the Peierls' transition but do not show evidence for spin-charge separation.  
The MDC peak widths at T = 30~K for the TCNQ spinon-band derived surface is $\sim$ 0.05b* \AA$^{-1}$ (Fig. 2(a)).  
This peak width implies a CDW coherence length of $\sim$ 120 \AA.  
The MDC peak width ($\sim$ 0.1b* \AA$^{-1}$, Fig. 2(b)) for the TTF-derived surface 
implies a shorter CDW coherence length of $\sim$ 60 \AA.  
These values are consistent with recent X-ray diffuse scattering studies \cite{Pouget04} 
and suggest that the TCNQ derived CDW order is more robust and induces the Peierls' order in the TTF stacks.  
The results indicate the 2$k_F$ and 4$k_F$ order in the TCNQ surface is derived from electron-electron correlations 
and the 2$k_F$ order in TTF surface alone is due to an electron phonon coupling induced Peierls' transition.  
This clarifies the 
\linebreak
 \begin{figure}
 \includegraphics[scale=.50]{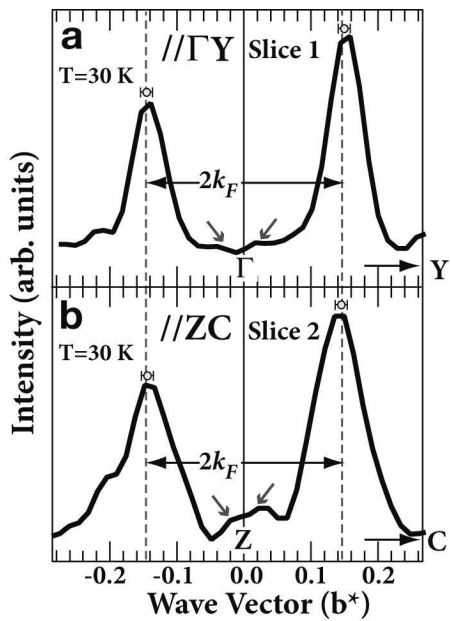}%
 \caption{\label{fig2}
 Momentum distributuion curves of TTF-TCNQ: 
 Fig. 2(a) and (b) show the momentum distribution curves (MDCs) 
 along slice 1 (TCNQ derived band) and slice 2 (TTF derived band) at T = 30~K obtained using ARPES spectra.  
 At T = 30~K, the separation in their positions (0.3 $\pm$ 0.01)b* matches the CDW vector (2$k_F$ = 0.295b*) 
 known from structural studies.  
 The peak widths imply a CDW coherence length of $\sim$ 120~\AA 
 for the TCNQ and $\sim$ 60~\AA for the TTF derived surfaces.  
 Small gray arrow marks the weak superstructure also seen in the maps (Fig. 1(d)).
 }
 \end{figure}
\noindent
specific role of TTF and TCNQ derived electronic states as a function of momentum 
and temperature compared with earlier work \cite{Tomkiewicz,Pouget,Shirane,Kagoshima,Takahashi}.

In addition to the main TTF and TCNQ Luttinger surfaces, additional weak features are seen at T = 30~K 
in the MDCs of Fig. 2(a) and (b), marked with small arrows.  
These features are better seen in Fig. 1(d), which is a gray scale image of Fig. 1(c), 
plotted in order to see the weak features with stronger contrast.  
The T = 30~K gray scale image shows many more features compared to the 200~K plot (Fig. 1(a)).  
We checked for the separation between the features along b* and find that the weak features occur exactly 
at multiples of 2$k_F$ and 4$k_F$ vectors obtained by a backfolding from the second Brillouin zone.  
The deviations from quasi 1D behavior in the weak features is attributed to the difference 
in peak widths and positions along b* in the main TTF and TCNQ Luttinger surfaces, 
which get multiplied for the weak features.  
A very weak feature is also seen in Fig. 1(a) near the $\Gamma$-point 
due to a backfolding of the 4$k_F$ CDW vector at T = 200~K.  
A schematic of the high- and low-T main Luttinger surfaces and the backfolded Luttinger surfaces is drawn 
in Fig. 1(e) and (f) to explain all the observed features due to the 2$k_F$ and 4$k_F$ CDW order 
and their multiples as seen in the maps.

The observation of temperature dependent Luttinger surfaces indicates 
that angle - resolved photoemission spectroscopy provide important insights about CDW order in a Luttinger liquid, 
and its spectral function retains Fermi liquid-like properties in terms of peaks in MDCs, 
although the quasiparticles in EDCs are suppressed \cite{Allen,Ren,Kim,Orgad,Gweon,Kivelson,Dzyaloshinskii,Zwick}.  
The evolution from a Mott-Insulator (MI) to Luttinger liquid and then on to a Fermi liquid 
with a concomitant transformation of the Fermi surface topology was predicted theoretically for the Bechgaard salts, 
in a model of correlated chains with interchain hybridization \cite{Biermann}.  
Optical measurements indeed reveal evidence for spin-charge separation in the Bechgaard salts 
and a dimensionally driven insulator-metal transition \cite{Vescoli}.  
Luttinger surface studies are eagerly awaited for this family of quasi 1-D correlated metals 
which also exhibit density wave order and superconductivity.  
If we identify the MI phase with the CDW ordered phase, 
the same theoretical phase diagram \cite{Biermann} holds for the topology change observed in TTF-TCNQ (Fig. 1(a)-(c)).  
While recent work has \cite{Ishii,Rauf} identified the power law characteristic of Luttinger liquids 
in angle-integrated photoemission spectra of Carbon nanotubes \cite{Ishii,Rauf}, 
and also a doping dependent Luttinger to Fermi liquid crossover \cite{Rauf}, 
ARPES of the Fermi/Luttinger surfaces in the Carbon nanotube systems are intrinsically difficult. 

In conclusion, the Luttinger surface topology, obtained from MDCs, changes from quasi-2D to quasi-1D with temperature.  
The high temperature quasi-2D surface exhibits 4$k_F$ CDW superstructure in the TCNQ derived holon band, 
in the absence of 2$k_F$ order.  
On decreasing temperature, a quasi-1D nested 2$k_F$ CDW order in the TCNQ spinon band and in the TTF surface is obtained.  
The results show the link in momentum-space between charge order and spin-charge separation in a Luttinger liquid.  
The present study indicates T-dependent Luttinger surface topology is very important 
for density-wave phenomena in low dimensional correlated systems.

\begin{acknowledgments}
This study was supported by a Grant-in-Aid for Scientific Research(No.16GS50219) 
from the Ministry of Education, Culture, Sports, Science and Technology of Japan.
\end{acknowledgments}

\end{document}